\documentclass[twocolumn,showpacs,preprintnumbers,amsmath,amssymb,
superscriptaddress]{revtex4-1}

\usepackage{graphicx}
\usepackage{dcolumn}
\usepackage{bm}

\begin{document}

%\preprint{APS/123-QED}

\title{Alpha-cluster structure in even-even nuclei around $^{94}$Mo}

\author{M.~A.~Souza}
\email{marsouza@if.usp.br} 
\affiliation{Instituto de F\'{\i}sica, Universidade de S\~{a}o Paulo, Rua do
Mat\~{a}o, Travessa R, 187, CEP 05508-090, Cidade Universit\'{a}ria, S\~{a}o
Paulo - SP, Brazil}
\affiliation{Instituto Federal de Educa\c{c}\~{a}o, Ci\^{e}ncia e Tecnologia de
S\~{a}o Paulo - Campus S\~{a}o Paulo, Rua Pedro Vicente, 625, CEP 01109-010,
Canind\'{e}, S\~{a}o Paulo - SP, Brazil}
\author{H.~Miyake}
\email{miyake@if.usp.br}
\affiliation{Instituto de F\'{\i}sica, Universidade de S\~{a}o Paulo, Rua do
Mat\~{a}o, Travessa R, 187, CEP 05508-090, Cidade Universit\'{a}ria, S\~{a}o
Paulo - SP, Brazil}

\date{\today}

\begin{abstract}
\begin{description}

\item[Background] The $\alpha$-cluster model was successful for describing
spectroscopic properties of light nuclei near the shell closures. Evidences
of the $\alpha$-cluster structure in $^{94}$Mo were shown, motivating the
search for the same structure in other intermediate mass nuclei.
\item[Purpose] The systematic analysis of the $\alpha$-cluster structure in
$^{94}$Mo and four even-even neighbouring nuclei.
\item[Method] The $\alpha$-cluster model with the local potential approach.
\item[Results] A good general description of the experimental ground state bands
is obtained with only one variable parameter. It is shown that the employed
$\alpha+\textrm{core}$ potential is weakly dependent on the angular momentum $L$
and such dependence may be described satisfactorily by a simple $L$-dependent
factor for the five nuclei. The radial parameter of the $\alpha+\textrm{core}$
potential reveals a linear trend in relation to the total nuclear radius and the
sum of the $\alpha$-cluster and core radii. The calculated $B(E2)$ transition
rates reproduce correctly the order of magnitude of almost all experimental data
without the use of effective charges. The rms intercluster separations and
reduced $\alpha$-widths obtained for the ground state bands point to a reduction
of the $\alpha$-cluster intensity with the increasing spin. Negative parity
bands are calculated and compared to available experimental levels.
The volume integral per nucleon pair and rms radii were calculated for the
studied $\alpha+\textrm{core}$ potentials, revealing a similarity with
the real parts of the optical potentials used to describe the $\alpha + ^{90}$Zr
and $\alpha + ^{92}$Mo elastic scattering at several energies.
\item[Conclusions] This work strongly indicates the presence of similar
$\alpha+\textrm{core}$ structures for the $N=52$ even-even nuclei of the Mo
region.

\end{description}
\end{abstract}

                            
%\keywords{cluster models, intermediate mass nuclei, $\alpha$-cluster structure,
%$^{90}$Sr, $^{92}$Zr, $^{94}$Mo, $^{96}$Ru, $^{98}$Pd}
                              
\maketitle

\section{Introduction}

The $\alpha$-cluster structure is an important aspect in the
analysis of nuclear spectral data. The cluster interpretation
represents a suitable way to describe nuclear states
which could be investigated by more complex microscopic treatments.
The $\alpha$-cluster model has been successful in reproducing
several levels of the energy spectra, electromagnetic properties, 
$\alpha$-emission widths and $\alpha$-particle elastic scattering data
in nuclei near the double shell closures.
The $\alpha$-cluster interpretation has often been used in studies on different
mass regions. In heavy and superheavy mass regions, recent studies use
this concept for analysis of the $\alpha$-decay phenomenon \cite{IA2014,QR2014,IW2014,DT2012}.
In light mass region, the $\alpha$-cluster interpretation is combined with various
approaches for describing properties of $n \alpha$ or $n \alpha + \textrm{\{other particles\}}$
nuclei; recent examples are the concept of nonlocalized clustering \cite{ZFH2013}, the
Hartree-Fock-Bogoliubov approach \cite{GS2013} and the quantum molecular dynamics
(or its antisymmetrized version) \cite{HMC2014,BCK2014}.

The earlier works of Buck, Dover and Vary \cite{BDV75} on $^{16}$O and $^{20}$Ne,
and Michel, Reidemeister and Ohkubo \cite{MRO88} on $^{44}$Ti, demonstrated the
success of the $\alpha$-cluster interpretation with the Local Potential Model (LPM)
to describe the first experimental energy bands
of the three nuclei. The model was also used to determine
the $B(E2)$ transition rates between the members of the calculated bands,
obtaining good agreement with the experimental data without the use of
effective charges. The calculated $\alpha$-decay widths for
$^{16}$O and $^{20}$Ne \cite{BDV75} provide a good reproduction
of the orders of magnitude of the respective experimental widths. The
subsequent work of Buck, Merchant and Perez \cite{BMP95}, which takes
a modified Woods-Saxon potential (W.S.$+$W.S.$^{3}$) for the
$\alpha + \textrm{core}$ interaction, confirmed the favorable
results for the nuclei $^{20}$Ne and $^{44}$Ti.

The study of the $\alpha$-cluster structure above double shell closures continued
in the heavier nuclei $^{94}$Mo and $^{212}$Po. Concerning $^{94}$Mo, Ref.~\cite{BMP95}
describes the $\alpha + ^{90}$Zr system through the nuclear W.S.$+$W.S.$^{3}$ potential,
where a set of fixed parameters is employed for nuclei of different mass regions.
The results yield a good description of the ground state band of
$^{94}$Mo and the correct order of magnitude of the experimental $B(E2)$ rates
without the use of effective charges \cite{BMP95}. The work of Ohkubo
\cite{O1995} analyses the same structure in $^{94}$Mo through a double folding
potential; in this case, a good description of the $\alpha + ^{90}$Zr elastic
scattering data is obtained, the $^{94}$Mo ground state band is reproduced
with a weak dependence on the quantum number $L$ and the absolute values of the
$B(E2)$ transition rates are well described with a small effective charge
($\delta e = 0.2e$). Following works of Michel \textit{et al.}~\cite{MRO2000}
and Souza and Miyake \cite{SM2005} reinforce the indications on the
$\alpha$-cluster structure in the $^{94}$Mo states.
Concerning $^{212}$Po, the results obtained in Refs.~\cite{BMP95,O1995,HMS1994,BJM1996}
for the ground state band and the $\alpha + ^{208}$Pb elastic scattering cross sections show
a similar level of agreement with the respective experimental data. Additionally, the model provides
a satisfactory description of the orders of magnitude of the experimental half-lives $T_{1/2}$
for $^{212}$Po.

The work of Mohr \cite{M2008} shows a study of the $\alpha + \textrm{core}$
structure in intermediate mass nuclei by means of a double folding nuclear potential.
Using the LPM, the calculations give favorable results for the $B(E2)$
rates of the ground state bands. However, an analysis of the intensity parameter
$\lambda$ of the nuclear potential shows a weak $L$ dependence and a significant
dependence on the proton number of the nucleus, mainly around the $Z=40$ subshell closure
and the $Z=50$ shell closure.

In this context, the purpose of
the present work is a detailed analysis of the $\alpha$-cluster structure in
$^{94}$Mo and the even-even neighbouring nuclei $^{90}$Sr, $^{92}$Zr, $^{96}$Ru
and $^{98}$Pd. The $\alpha + \textrm{core}$ systems are calculated from the
viewpoint of the LPM. Several properties of the ground state bands
are discussed and indications of $\alpha + \textrm{core}$ negative parity bands
are sought. Possible structure patterns in these nuclei are investigated.

\section{Selection of preferential nuclei for $\alpha$-clustering}

A preliminary question is the choice of a criterion to determine the most
favourable nuclei for $\alpha $-clustering in a specified mass region. An usual
procedure is the selection of nuclei with the configuration of $\alpha$-cluster
plus doubly (or even singly) closed shell core. However, such procedure does not
allow for a systematic analysis of any set of nuclei. In this work, we use a
criterion based only on experimental data of binding energy \cite{AWT2003}. An
appropriate quantity for comparing different nuclei of a set is the variation of
average binding energy per nucleon of the system due to the
$\alpha + \mathrm{core}$ decomposition. This value is given by

\begin{equation}
\frac {Q_\alpha}{A_T}=\frac{B_\alpha +B_{\mathrm{core}}-B_T}{A_T}\;, 
\end{equation}

\noindent where $Q_\alpha$ is the $Q$-value for $\alpha$-separation, $A_T$ is
the mass number of the total nucleus and $B_\alpha $, $B_{\mathrm{core}}$ and
$B_T$ are the experimental binding energies of the $\alpha $-cluster, the core
and the total nucleus, respectively. Thus, an absolute (or local) maximum of
$Q_{\alpha}/A_T$ indicates the preferred nucleus for $\alpha $-clustering in
comparison with the rest of (or neighbouring) nuclei in the set.

\begin{figure*}
\includegraphics[scale=1.3]{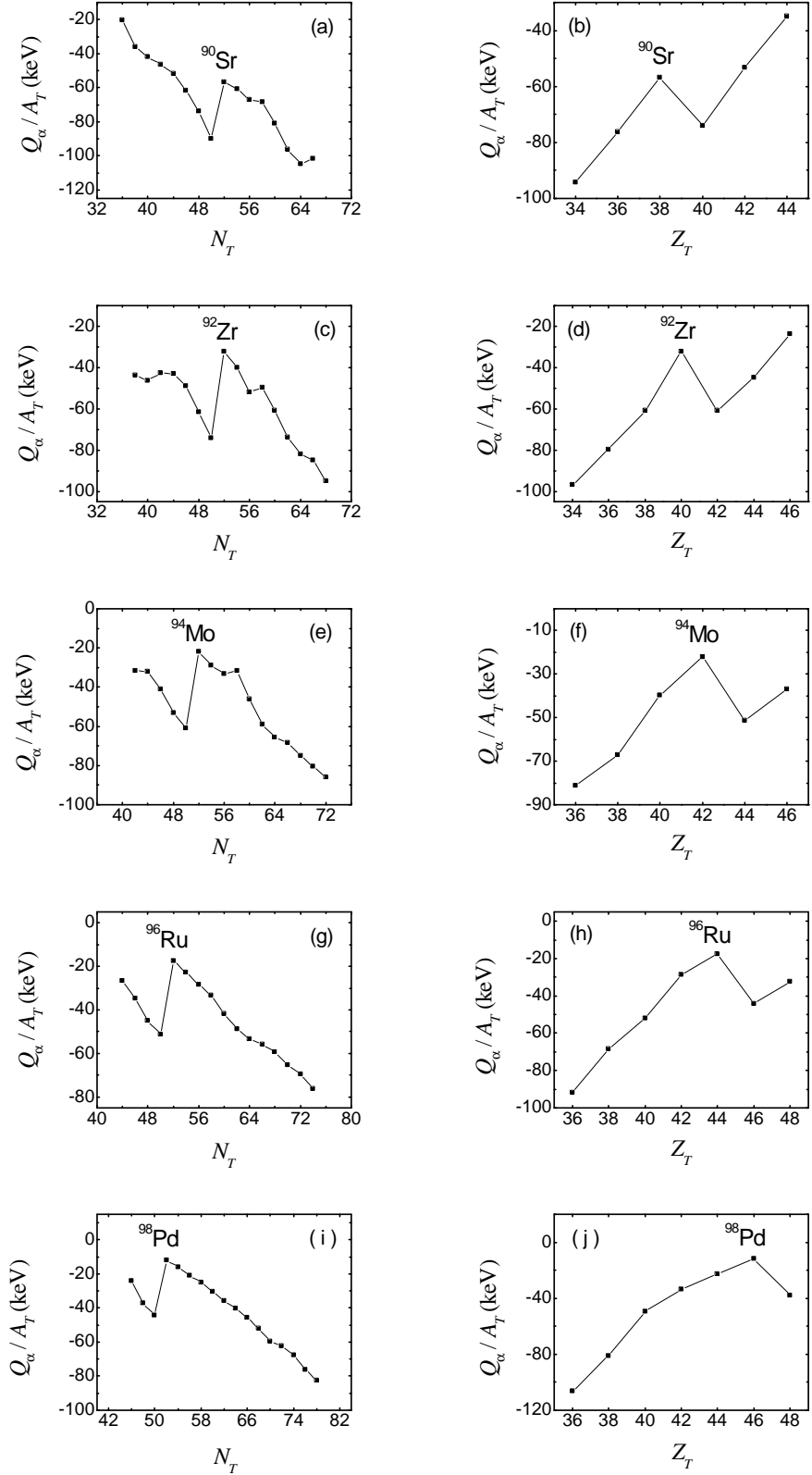}
\caption{$Q_{\alpha}/A_T$ values obtained for the $\alpha $ + core
decomposition of even-even Sr (a), Zr (c), Mo (e), Ru (g) and Pd (i)
isotopes as a function of the total neutron number $N_{T}$, and even-even
$A=90$ (b), $A=92$ (d), $A=94$ (f), $A=96$ (h) and $A=98$ (j) isobars as a
function of the total charge number $Z_{T}$. The $Q_{\alpha}/A_T$ peaks
corresponding to $^{90}$Sr, $^{92}$Zr, $^{94}$Mo, $^{96}$Ru and $^{98}$Pd
are indicated.}
\label{Figure_Q_alpha}
\end{figure*}

Restricted atomic number and mass number regions are defined around $^{94}$Mo
for the $\alpha $-cluster analysis. The $Q_{\alpha}/A_T$ value is used for
comparison of even-even isotopes between $Z=38$ and $Z=46$, and even-even
isobars between $A=90$ and $A=98$. FIG.~\ref{Figure_Q_alpha} shows graphically
the $Q_{\alpha}/A_T$ values for different sets of nuclei. Observing the sets of
even-even isotopes, a local $Q_{\alpha}/A_T$ peak is seen for $^{90}$Sr and
absolute $Q_{\alpha}/A_T$ peaks are seen for $^{92}$Zr, $^{94}$Mo, $^{96}$Ru and
$^{98}$Pd. Observing the sets of even-even isobars, local $Q_{\alpha}/A_T$ peaks
are seen for $^{90}$Sr and $^{92}$Zr, and absolute $Q_{\alpha}/A_T$ peaks are
seen for $^{94}$Mo, $^{96}$Ru and $^{98}$Pd. Such results imply that $^{90}$Sr,
$^{92}$Zr, $^{94}$Mo, $^{96}$Ru and $^{98}$Pd are preferential nuclei for
$\alpha$-clustering if they are compared with the respective even-even isotopes
and isobars simultaneously.

It is important to note that the five mentioned nuclei have the same number of
neutrons ($N = 52$), that is, their cores have the same magic number $N = 50$.
Therefore, the $Q_{\alpha}/A_T$ analysis provides a strong indication that the
number $N_{\mathrm{core}} = 50$ characterizes the most favourable nuclei for
$\alpha$-clustering in the Mo mass region.

\begin{figure}
\includegraphics[scale=0.35]{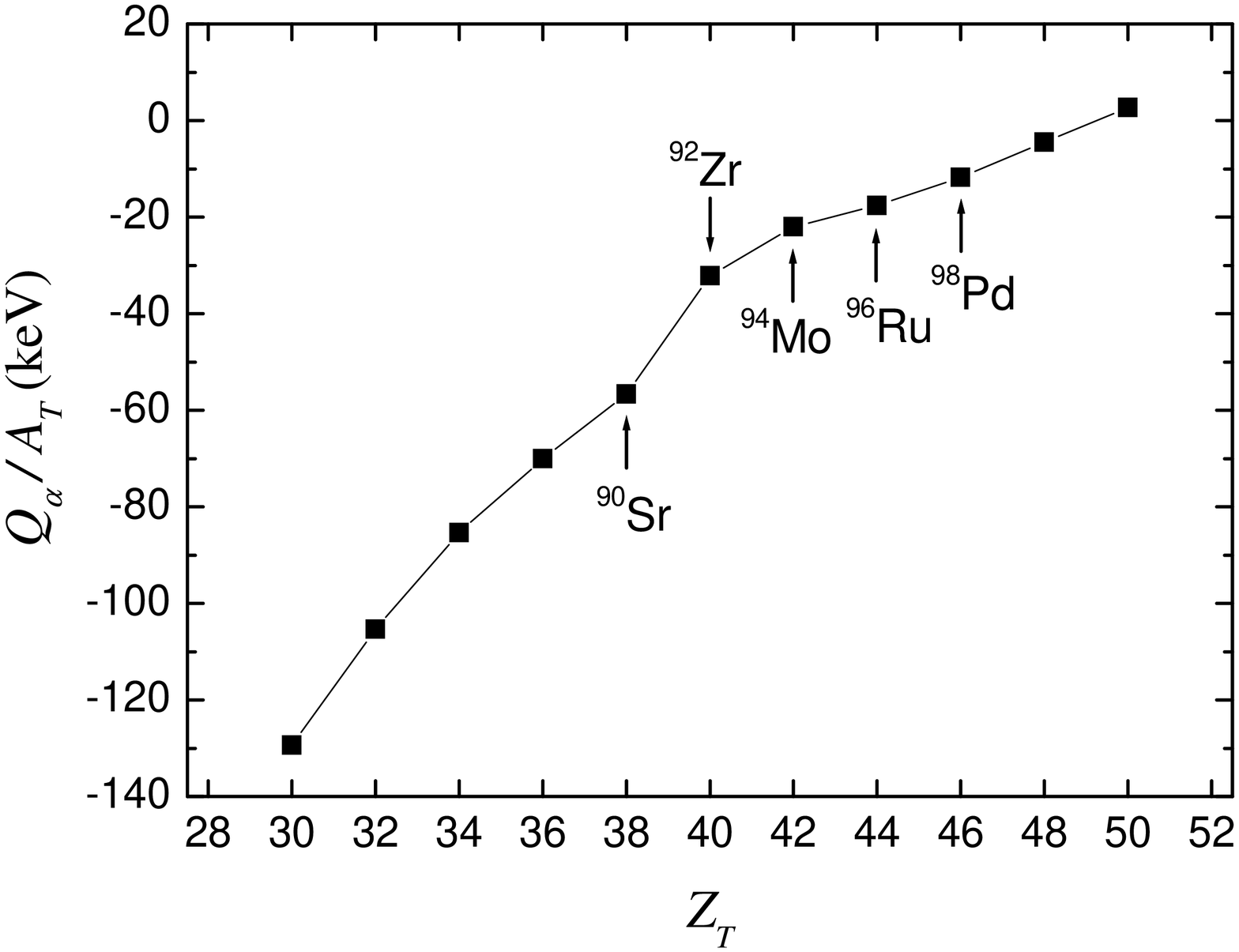}
\caption{$Q_{\alpha}/A_T$ values obtained for the $\alpha $ + core
decomposition of even-even $N=52$ isotones as a function of the total charge
number $Z_{T}$. The dots corresponding to $^{90}$Sr, $^{92}$Zr, $^{94}$Mo,
$^{96}$Ru and $^{98}$Pd are indicated.}
\label{Figure_Q_alpha_N=52}
\end{figure}

A comparison of the $Q_{\alpha}/A_T$ values for
the set of even-even $N=52$ isotones is presented in
FIG.~\ref{Figure_Q_alpha_N=52}. It is noted that the $Q_{\alpha}/A_T$
variation rate increases moderately between $Z_{T}=38$ and $Z_{T}=40$ in
comparison with other neighbouring nuclei. Therefore, the effect of the
subshell closure at $Z_{\mathrm{core}} = 38$ is more pronounced than at
$Z_{\mathrm{core}} = 40$ for the increase of $Q_{\alpha}/A_T$.
The fact that there are no abrupt $Q_{\alpha}/A_T$ variations around $^{94}$Mo 
suggests that this nucleus and its neighbouring isotones should present 
similar $\alpha + \textrm{core}$ structures. Indeed, this indication is
verified by different properties discussed in Section \ref{Sec:Results}.

\section{The $\alpha $-cluster model}
\label{Section3}

The properties of the nucleus are viewed in terms of a preformed
$\alpha $-particle orbiting an inert core. Internal excitations of the
$\alpha$-cluster and the core are not considered in the calculations. The
$\alpha + \mathrm{core}$ interaction is described through a local
phenomenological potential \mbox{$V(r)=V_C(r)+V_N(r)$} containing the Coulomb
and nuclear terms. For the nuclear potential, we adopt the form

\begin{eqnarray}
\lefteqn{ V_N(r)=-V_0\left\{ \frac b{1+\exp [(r-R)/a]} \right. {} }
                                                          \nonumber\\[4pt]
 & \qquad & {} \left. + \frac{1-b}{\{1+\exp[(r-R)/3a]\}^3}\right\} 
\label{eq:Nuc_Pot}
\end{eqnarray}

\noindent proposed by Buck, Merchant and Perez \cite{BMP95}. The Coulomb
potential $V_C(r)$ is taken to be that of an uniform spherical charge
distribution of radius $R_C=R$. The inclusion of the centrifugal term
results in the effective potential

\begin{equation}
V_{\mathrm{eff}}(r)=V(r)+\frac{L\left(L+1\right)\hbar^{2}}{2\mu r^{2}}\;,
\label{eq:Veff}
\end{equation}

\noindent where $\mu$ is the reduced mass of the $\alpha + \mathrm{core}$
system.

The general description of the ground state
bands is made with the fixed parameters $V_0 = 220$ MeV, $a = 0.65$ fm and
$b = 0.3$, while $R$ is fitted separately for each nucleus. The values of $V_0$,
$a$ and $b$, which are obtained from Ref.~\cite{BMP95}, have been fitted to
reproduce satisfactorily the experimental excitation energies of the ground
state bands of $^{20}$Ne, $^{44}$Ti, $^{94}$Mo and $^{212}$Po, as well as the
experimental $\alpha $-decay half-lives for several even-even heavy nuclei.
Using a procedure similar to that employed in Ref.~\cite{BMP95},
the radial parameter $R$ is fitted in the present work to reproduce the
experimental 4$^{+}$ member of the ground state band of each nucleus. The
obtained values for $R$ are shown in TABLE \ref{Tab_Radii}.

\begin{table}
\caption{Values of the radial parameter $R$ for the intermediate mass nuclei
analysed in this work.}
\label{Tab_Radii}
\begin{ruledtabular}
\begin{tabular}{ccc}
Nucleus & System & $R$ (fm)\\[2pt] \hline
&  &  \\[-6pt] 
$^{90}$Sr & $\alpha+{}^{86}$Kr & 5.321 \\
$^{92}$Zr & $\alpha+{}^{88}$Sr & 5.295 \\
$^{94}$Mo & $\alpha+{}^{90}$Zr & 5.784 \\
$^{96}$Ru & $\alpha+{}^{92}$Mo & 5.808 \\
$^{98}$Pd & $\alpha+{}^{94}$Ru & 5.825 \\
\end{tabular}
\end{ruledtabular}
\end{table}

The Pauli principle requirements for the $\alpha$ valence nucleons are
introduced through the quantum number

\begin{equation}
G = 2N + L \;,
\end{equation}

\noindent where $N$ is the number of internal nodes in
the radial wave function and $L$ is the orbital angular momentum.
The global quantum number $G$ identifies the bands of states. In this way, the
restriction $G\geq G_{\mathrm{g.s.}}$ is applied, where $G_{\mathrm{g.s.}}$
corresponds to the ground state band. The value $G_{\mathrm{g.s.}}=14$ is
employed for $^{90}$Sr and $^{92}$Zr, and $G_{\mathrm{g.s.}}=16$ for
$^{94}$Mo, $^{96}$Ru and $^{98}$Pd. These two values are obtained from the
Wildermuth condition \cite{WT1977} considering the $(pf)^2(sdg)^2$ and $(sdg)^4$
configurations, respectively.

The energy levels and associated radial wave functions are calculated
by solving the Schr\"{o}dinger radial equation for the reduced mass of the
$\alpha + \mathrm{core}$ system.

\section{Results}
\label{Sec:Results}

\subsection{The parameter $R$}
\label{Subsec:R}

\begin{figure}
\includegraphics[scale=0.7]{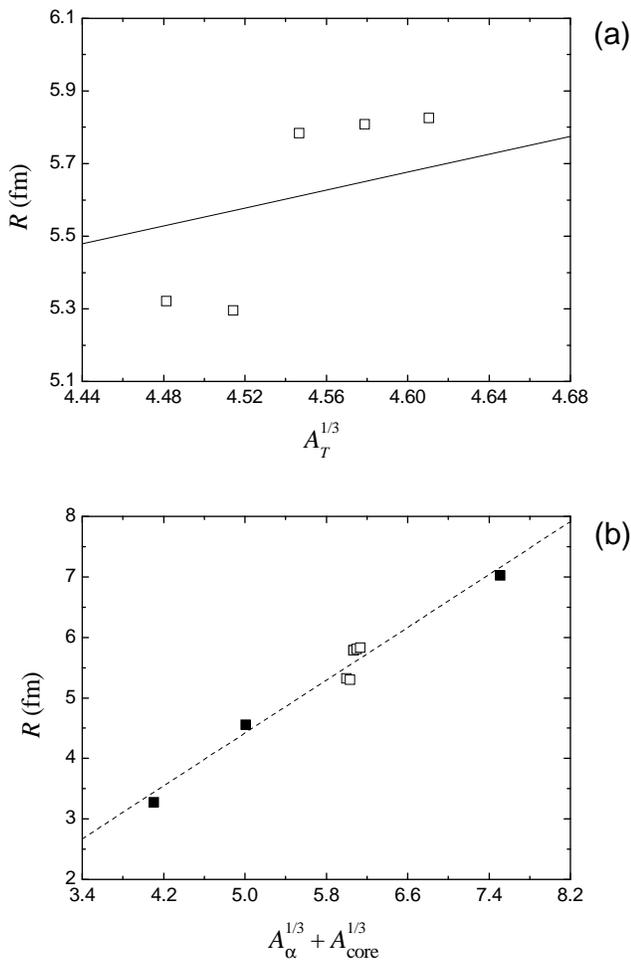}
\caption{Values of the parameter $R$ for the nuclei $^{90}$Sr, $^{92}$Zr,
$^{94}$Mo, $^{96}$Ru and $^{98}$Pd (open squares) as a function of $A_{T}^{1/3}$
(graph \textbf{(a)}) and $A_{\alpha}^{1/3} + A_{\mathrm{core}}^{1/3}$
(graph \textbf{(b)}). For a more comprehensive analysis, the $R$ values for
$^{20}$Ne, $^{44}$Ti and $^{212}$Po (full squares) are also shown in graph
\textbf{(b)}. The full line shows eq.~\eqref{eq:Fit_R_1} fitted to describe the
$R$ values of the five intermediate mass nuclei. The dashed line shows
eq.~\eqref{eq:Fit_R_2} fitted to describe the $R$ values of $^{20}$Ne,
$^{44}$Ti, $^{94}$Mo and $^{212}$Po.}
\label{Figure_R}
\end{figure}

TABLE \ref{Tab_Radii} presents an increase of the parameter $R$
with the nuclear mass number, except to a decrease from $^{90}$Sr to $^{92}$Zr.
Such feature motivates the search for a relation of $R$ to the nuclear radius.
FIG.~\ref{Figure_R}(a) shows $R$ as a function of $A_{T}^{1/3}$ and
FIG.~\ref{Figure_R}(b) as a function of
$A_{\alpha}^{1/3} + A_{\mathrm{core}}^{1/3}$, where $A_{T}$, $A_{\alpha}$
and $A_{\mathrm{core}}$ are the mass numbers of the total nucleus, the
$\alpha$-particle and the core, respectively. FIG.~\ref{Figure_R}(a) confirms
the increasing trend of $R$ with $A_{T}^{1/3}$, but a more pronounced increase
of $\approx$ 0.5 fm occurs between $^{92}$Zr and $^{94}$Mo. This is an effect of
the quantum number $G_{\mathrm{g.s.}}$ applied for each nucleus, since the use
of a lower (or higher) number for $G_{\mathrm{g.s.}}$ results in a lower (or
higher) value for the radial parameter of the $\alpha + \mathrm{core}$
potential, considering the other parameters as fixed.

For a more comprehensive analysis, the $R$ values for $^{20}$Ne, $^{44}$Ti and
$^{212}$Po have been determined (see TABLE \ref{Tab_Radii_2}) with the same
procedures described in Section \ref{Section3}, except to the choice of
$G_{\mathrm{g.s.}} = 20$ for $^{212}$Po, which is cited in Ref.~\cite{BMP95} as
a more appropriate number to describe the experimental spectrum of $^{212}$Po in
comparison with the quantum number predicted by the Wildermuth condition.

We have made linear fits with one and two parameter(s) for $R$ as a function of
$A_{T}^{1/3}$ and $R$ as a function of
$A_{\alpha}^{1/3} + A_{\mathrm{core}}^{1/3}$. Two sets of nuclei
have been considered in the fits: the set of five intermediate mass nuclei and
the set of nuclei with $\alpha$-clustering above double shell closure
($^{20}$Ne, $^{44}$Ti, $^{94}$Mo and $^{212}$Po). Comparing the fitted
functions, it was verified that the relation

\begin{equation}
R = 1.234\, A_{T}^{1/3}\,\mathrm{(fm)}
\label{eq:Fit_R_1}
\end{equation}

\noindent (see FIG.~\ref{Figure_R}(a)) provides a good description of the $R$
values for the five intermediate mass nuclei and also for $^{20}$Ne, $^{44}$Ti
and $^{212}$Po. However, the best fit obtained for the set \{$^{20}$Ne,
$^{44}$Ti, $^{94}$Mo, $^{212}$Po\} is given by

\begin{equation}
R = 1.092(A_{\alpha}^{1/3} + A_{\mathrm{core}}^{1/3}) - 1.041\,\mathrm{(fm)}\;,
\label{eq:Fit_R_2}
\end{equation}

\noindent which is shown graphically in FIG.~\ref{Figure_R}(b). The correlation
coefficient for the points $(R,A_{T}^{1/3})$ of the five intermediate mass
nuclei is $r \approx 0.882$, and the same measure for the points
$(R,(A_{\alpha}^{1/3} + A_{\mathrm{core}}^{1/3}))$ of the set \{$^{20}$Ne,
$^{44}$Ti, $^{94}$Mo, $^{212}$Po\} is $r \approx 0.993$. These results
confirm the linear trend of the parameter $R$ in relation to the total nuclear
radius and the sum of the $\alpha$-cluster and core radii.

\begin{table}
\caption{Values of the radial parameter $R$ and the quantum number
$G_{\mathrm{g.s.}}$ for $^{20}$Ne, $^{44}$Ti and $^{212}$Po.}
\label{Tab_Radii_2}
\begin{ruledtabular}
\begin{tabular}{cccc}
Nucleus & System & $R$ (fm) & $G_{\mathrm{g.s.}}$\\[2pt] \hline
&  &  &  \\[-6pt] 
$^{20}$Ne & $\alpha+{}^{16}$O & 3.272 & 8 \\
$^{44}$Ti & $\alpha+{}^{40}$Ca & 4.551 & 12 \\
$^{212}$Po & $\alpha+{}^{208}$Pb & 7.019 & 20 \\
\end{tabular}
\end{ruledtabular}
\end{table}

\subsection{Ground state bands}

\begin{figure*}
\includegraphics[scale=0.9]{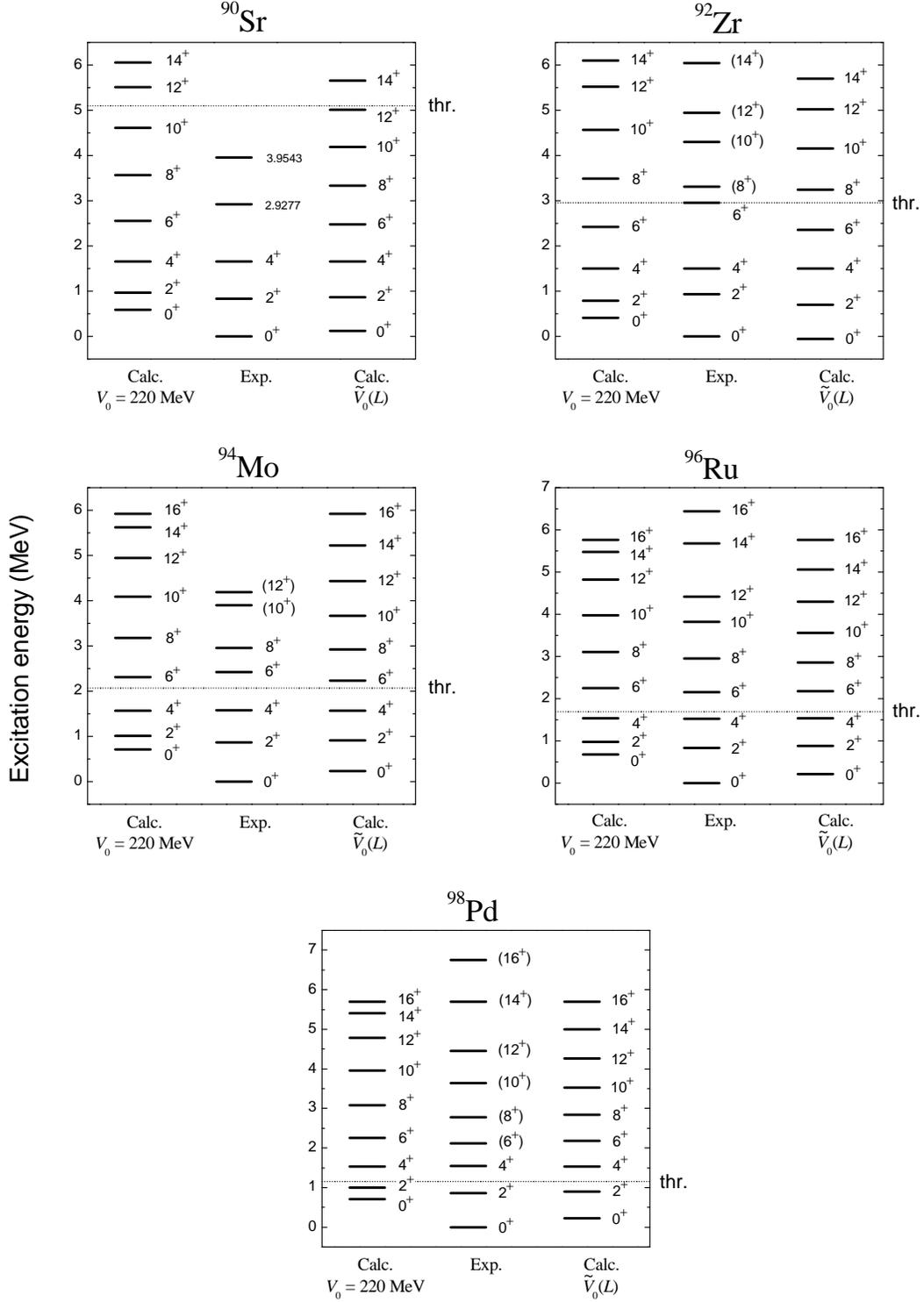}
\caption{Calculated $\alpha + \textrm{core}$ energies for the ground state
bands of $^{90}$Sr ($G=14$), $^{92}$Zr ($G=14$), $^{94}$Mo ($G=16$), $^{96}$Ru
($G=16$) and $^{98}$Pd ($G=16$) in comparison with the available experimental
excitation energies. The label \mbox{{\bf Calc.~$V_0 = 220$ MeV}} indicates the
calculated g.s.~bands with the fixed depth parameter, and the label
{\bf Calc.~$\tilde{V}_0(L)$} indicates the calculated g.s.~bands with the
$L$-dependent depth parameter (see the text for details of $\tilde{V}_0(L)$).
The dotted lines indicate the $\alpha + \textrm{core}$ thresholds. The
experimental data are from Refs.~\cite{B1997,B2012,AS2006,AS2008,SH2003}.}
\label{Figure_Espec_Pos}
\end{figure*}

Using the $\alpha$-core potential described in Section \ref{Section3},
we have calculated the ground state bands for $^{90}$Sr, $^{92}$Zr,
$^{94}$Mo, $^{96}$Ru and $^{98}$Pd. The obtained levels are compared with the
corresponding experimental energies in FIG.~\ref{Figure_Espec_Pos} (see
the label Calc.~$V_0 = 220$ MeV). The theoretical bands of the five nuclei give
a good overall description of the experimental spectra, if we consider that
only the parameter $R$ is varied for each nucleus (see \mbox{TABLE \ref{Tab_Radii}}) and
the fixed parameters $V_0$, $a$ and $b$ have been adjusted \cite{BMP95} to reproduce
spectroscopic properties of nuclei of different mass regions.
There are experimental levels of $^{92}$Zr, $^{94}$Mo and $^{98}$Pd with
uncertain assignments and the definitive identification of such states
is desirable. The experimental levels $E_x = 2.9277$ MeV and $E_x = 3.9543$ MeV
of $^{90}$Sr do not present identified spins and parities, however, we
suggest an association with the theoretical states 6$^{+}$ and 8$^{+}$,
respectively, since these levels and the experimental 4$^{+}$ state
($E_x = 1.65591$ MeV) are connected by intense $\gamma$-transitions
\cite{B1997}.

The $\alpha + \textrm{core}$ potential produces a quasirotational behaviour for
the first states of the calculated ground state bands, resulting in a rougher
description of the spacing between the 0$^{+}$ bandhead and the 2$^{+}$ state
for the five nuclei. It must to be mentioned that this feature is observed
in nuclei of different mass regions with the same $\alpha + \textrm{core}$
potential form \cite{BMP95} and other $\alpha + \textrm{core}$ potential forms
\cite{BDV75,MRO88,MRO2000}. The variation of spacing between the calculated
intermediate states is small, while the highest spin states show a compression.
Such behaviour is typical for the nuclear W.S.$+$W.S.$^{3}$ potential applied
at the intermediate and heavy mass regions, where the calculated bands present
high spin members. The difficulty to reproduce simultaneously the spacing between
the 0$^{+}$ and 2$^{+}$ states and higher spin states above 2$^{+}$ shows the
convenience of choosing the experimental 4$^{+}$ state for the fit of the radial
parameter $R$.

\begin{figure}
\includegraphics[scale=0.38]{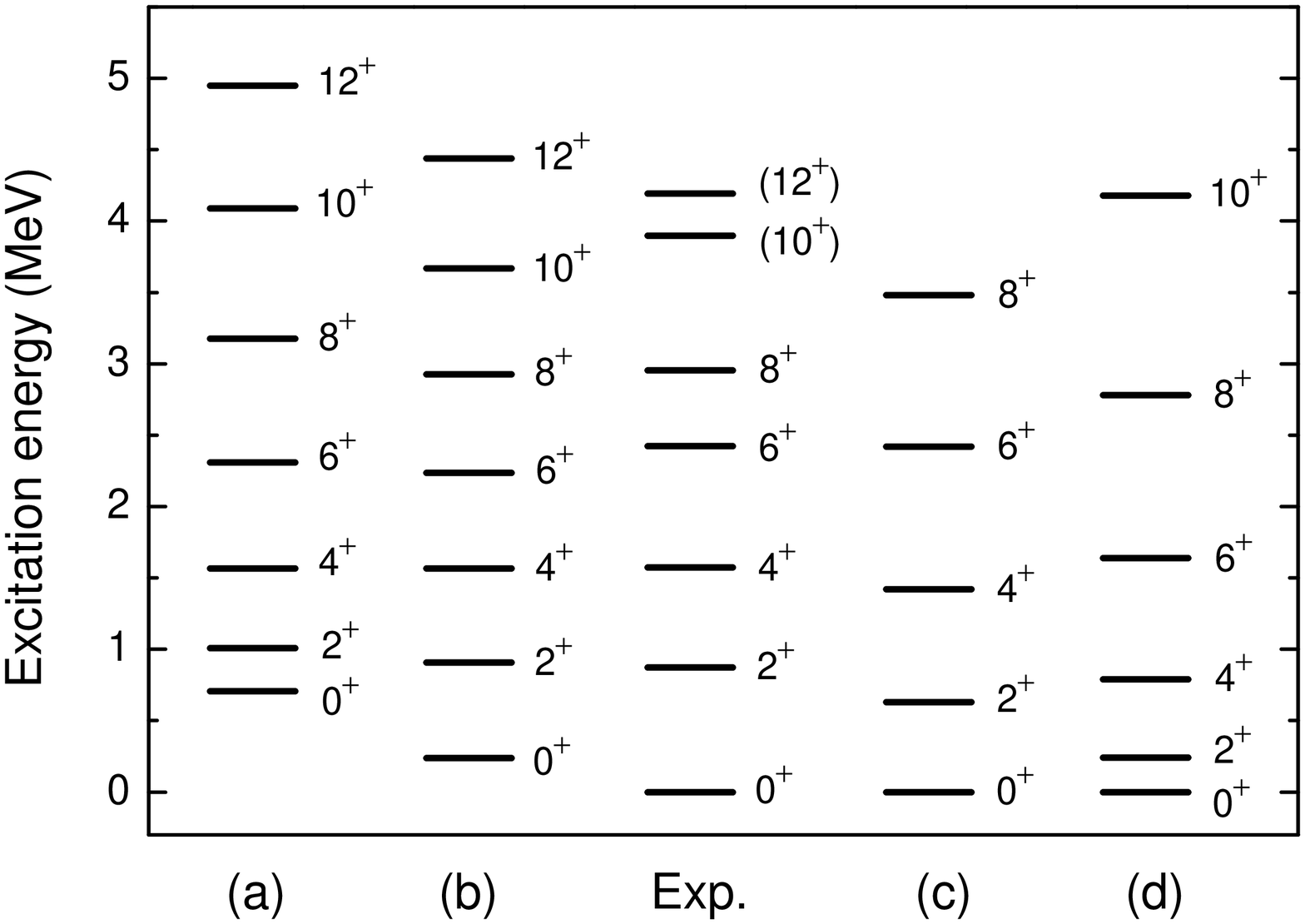}
\caption{Comparison of different calculations for the ground state band ($G=16$)
of $^{94}$Mo with the experimental excitation energies. Calc.~{\bf(a)} is
obtained in this work with the W.S.$+$W.S.$^{3}$ potential and fixed
depth $V_0 = 220$ MeV. Calc.~{\bf(b)} is obtained in this work with the
W.S.$+$W.S.$^{3}$ potential and $L$-dependent depth parameter (details in
the text). Calc.~{\bf(c)} \cite{O1995,MOR1998} is obtained with a double
folding potential using a DDM3Y nucleon-nucleon interaction and a
$L$-dependent normalization factor. Calc.~{\bf(d)} \cite{MRO2000}
is obtained with a phenomenological potential of shape
$(1 + \mathrm{Gaussian})\times \mathrm{W.S.}^{2}$. The potentials of {\bf(c)}
and {\bf(d)} were also used to reproduce $\alpha + ^{90}$Zr elastic
scattering data.}
\label{Figure_Comp_94Mo}
\end{figure}

The W.S.$+$W.S.$^{3}$ potential with fixed parameters yields better results than 
other potential forms used for intermediate mass nuclei. E.g., it is known
that the simple Woods-Saxon potential generates an inverted spectrum 
where the $G^{+}$ state has the lowest energy and $0^{+}$ state has the
highest energy \cite{M2008}. The folding potential produces a band with
rotational feature, but excessively compressed, and the inclusion of
an $L$-dependent strength parameter is required \cite{M2008,MOR1998}.
The $\alpha + \textrm{core}$ potential of shape
$(1 + \mathrm{Gaussian})\times \mathrm{W.S.}^{2}$ used for $^{94}$Mo \cite{MRO2000} 
produces a rotational type band (see FIG.~\ref{Figure_Comp_94Mo}), which 
is incompatible with the respective experimental spectrum.
It is observed in FIG.~\ref{Figure_Espec_Pos} that the experimental spectra
of other $N = 52$ nuclei are also quite different from the pure
rotational spacing. The W.S.$+$W.S.$^{3}$ potential with fixed parameters is shown
to be favorable mainly to describe the intermediate levels above the
$2^{+}$ state, since the rotational behavior of its calculated band attenuates
significantly above this level. 

Analysing the calculated positive parity states above the
$\alpha + \textrm{core}$ threshold, it is verified that the extension of the
effective potential barrier (see eq.~\eqref{eq:Veff}) varies within the range
$204 \; \mathrm{fm} < \Delta r < 476 \; \mathrm{fm}$ for the resonant states
nearest to the threshold, and within the range
$31 \; \mathrm{fm} < \Delta r < 114 \; \mathrm{fm}$ for the highest spin states
($14^{+}$ for $^{90}$Sr and $^{92}$Zr and $16^{+}$ for $^{94}$Mo, $^{96}$Ru and
$^{98}$Pd). The height of the effective potential barrier in relation to the
resonant energies varies within the range
$13 \; \mathrm{MeV} < \Delta E < 25 \; \mathrm{MeV}$ for the resonant states
nearest to the threshold, and within the range
$27 \; \mathrm{MeV} < \Delta E < 31 \; \mathrm{MeV}$ for the highest spin
states. These features demonstrate that the positive parity states above the
threshold are strongly bound and their resonant behaviour is negligible.

Several members of the ground state bands of $^{92}$Zr and $^{94}$Mo are
identified in $\alpha$-transfer experiments. The $^{92}$Zr states from 0$^{+}$
to (12$^{+}$) are populated in the \mbox{$^{88}$Sr($^7$Li,$2np\gamma$)$^{92}$Zr}
reaction \cite{B2012,B1976} and the $^{94}$Mo states from 0$^{+}$
to 6$^{+}$ are populated in the $^{90}$Zr($^{16}$O,$^{12}$C$\gamma$)$^{94}$Mo
reaction \cite{AS2006,BDM1972}. Such experimental information reinforce the
choice of the mentioned states for comparison with the calculated bands.
The work of Yamaya {\it et al.}~\cite{YKF1998} shows the spectrum of $^{94}$Mo
measured by the \mbox{$^{90}$Zr($^6$Li,{\it d})$^{94}$Mo} reaction, suggesting
that the 0$^{+}$, 2$^{+}$, 4$^{+}$ and 6$^{+}$ members of the ground state
band of $^{94}$Mo are populated in this reaction; however, the same work
regards this experiment as unsuccessful because of the strong effect of the
$^{12}$C and $^{16}$O contaminants on the target, which makes the
identification of the energy levels very difficult.
The main tables of spectroscopic data of $^{90}$Sr \cite{B1997}, $^{96}$Ru
\cite{AS2008} and $^{98}$Pd \cite{SH2003} do not present information from
$\alpha$-transfer reactions and new experiments will be useful to verify the
experimental levels selected for these three nuclei.

The radial wave functions of the positive parity states have been determined
to investigate other properties. A bound state approximation has
been used to determine the radial wave functions of the resonant states.
For the calculation of the functions, the depth $V_0$ is adjusted for each state
in order to reproduce the experimental excitation energies of the five nuclei.
Observing the values of $V_0$ in TABLE \ref{Table1}, we note a very small
relative variation in comparison with the fixed value (220 MeV) used for the
first calculation of the ground state bands. This fact confirms that the fixed
depth $V_0 = 220$ MeV is appropriate for an overall description of the spectra
and shows that $V_0$ is weakly dependent on $L$.

\begin{figure}
\includegraphics[scale=0.49]{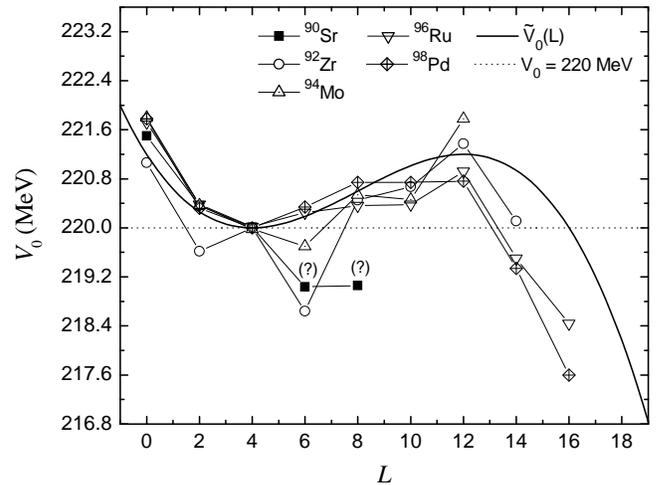}
\caption{Depths $V_{0}$ fitted to reproduce the experimental energy levels of
the g.s.~bands of $^{90}$Sr, $^{92}$Zr, $^{94}$Mo, $^{96}$Ru and $^{98}$Pd as
a function of the angular momentum $L$. The solid line shows the function
$\tilde{V}_0(L)$ (see eq.~\eqref{eq:FuncPol_V0}) chosen for a general
description of the graphs $V_{0} \times L$. The dotted line indicates the
constant value $V_{0} =$ 220 MeV used for a first description of the spectra.
Note that the scale for $V_0$ shows only very minor variations of approx.~$\pm$ 1.5 \%
from $V_{0} =$ 220 MeV. The dots accompanied by the symbol {\bf(?)} are related
to the experimental $^{90}$Sr levels with unidentified spins and parities, and an
association with the corresponding $L$ numbers is suggested (details in the text and
FIG.~\ref{Figure_Espec_Pos}).}
\label{Figure_V0}
\end{figure}

The fitted values of $V_0$ are shown graphically in FIG.~\ref{Figure_V0} as a
function of $L$. It is noted that, with few exceptions, the variation of $V_0$
is similar for the five nuclei. There are local minima of $V_0$ at $L=4$ and
$L=6$, and local maxima at $L=12$. Such features are favourable to describe
approximately the behaviour of $V_0$ by a function $\tilde{V}_0(L)$. For this
purpose, we have chosen a third degree polynomial function which is determined
by two previously selected points $(L,\tilde{V}_0)$:

\[
(L_{\mathrm{min}},\tilde{V}_0^{\mathrm{min}})=(4,220)\;\mathrm{and}\;
(L_{\mathrm{max}},\tilde{V}_0^{\mathrm{max}})=(12,221.2)\;.
\label{eq:Valor_max_min}
\]

\noindent $\tilde{V}_0^{\mathrm{min}}$ and $\tilde{V}_0^{\mathrm{max}}$ are
given in MeV. The points $(L_{\mathrm{min}},\tilde{V}_0^{\mathrm{min}})$ and
$(L_{\mathrm{max}},\tilde{V}_0^{\mathrm{max}})$ are the local minimum and
maximum of the function $\tilde{V}_0(L)$, respectively. The values of
$L_{\mathrm{min}}$ and $\tilde{V}_0^{\mathrm{min}}$ are appropriate to reproduce
the local minima of $V_0$ for $^{96}$Ru and $^{98}$Pd and also the decreasing
behaviour of the graphs $V_0 \times L$ between $L=0$ and $L=4$ . The values of
$L_{\mathrm{max}}$ and $\tilde{V}_0^{\mathrm{max}}$ are appropriate to reproduce
approximately the local maxima of $V_0$ for $^{92}$Zr, $^{94}$Mo, $^{96}$Ru and
$^{98}$Pd. Thus one obtains

\begin{equation}
\tilde{V}_0(L)=c_{1}(L-L_{\mathrm{min}})^{3}+c_{2}(L-L_{\mathrm{min}})^{2}+
\tilde{V}_0^{\mathrm{min}}\;,
\label{eq:FuncPol_V0}
\end{equation}

\noindent where

\[
c_{1}=-4.6875\times10^{-3}\,\mathrm{MeV} \quad\mathrm{and}\quad c_{2}=0.05625\,
\mathrm{MeV}\;.
\]

Eq.~\eqref{eq:FuncPol_V0} is shown graphically in FIG.~\ref{Figure_V0}. Between
$L=0$ and $L=12$, the function $\tilde{V}_{0}(L)$ provides a good general
description of the graphs $V_{0} \times L$ and a better approximation for most
of the $V_0$ values in comparison with the constant value $V_0 = 220$ MeV. For
$L > 12$, the function gives a rough description of the decreasing behaviour of
the graphs $V_{0} \times L$.

The function $\tilde{V}_{0}(L)$ has been used for a second calculation of the
ground state bands. The nuclear potential form of eq.~\eqref{eq:Nuc_Pot} is used
with $\tilde{V}_{0}(L)$ as the $L$-dependent depth factor. The results are
compared with the experimental energies in FIG.~\ref{Figure_Espec_Pos} (see
label Calc.~$\tilde{V}_{0}(L)$). With the $L$-dependent depth, the variation of
spacing between the calculated levels is very small. Such feature contributes to
a better description of the $0^{+}$ bandheads of the five nuclei and the $2^{+}$
states of $^{90}$Sr, $^{94}$Mo, $^{96}$Ru and $^{98}$Pd in comparison with the
first calculation where $V_0$ is fixed. Between $L = 6$ and $L = 12$, most of
the states are better described, however, the improvement in comparison with the
fixed $V_0$ calculation is small. For $L = 14$ and $L = 16$, the $L$-dependent
depth do not contribute for a better description of the experimental levels, but
avoids the band compression effect for the higher spin states that is
incompatible with the experimental spectra of $^{92}$Zr, $^{96}$Ru and
$^{98}$Pd.

\begin{table*}
\caption{Calculated values for the rms intercluster separation
($\langle R^2 \rangle ^{1/2}$), the ratio of $\langle R^2 \rangle ^{1/2}$ to
the sum of the experimental rms charge radii of the $\alpha$-cluster and the
core (\,$\langle R^2\rangle ^{1/2}/(\langle r^2\rangle _\alpha ^{1/2}+
\langle r^2\rangle _{\mathrm{core}}^{1/2})$\,), the reduced $\alpha $-width
($\gamma _\alpha ^2$) and the dimensionless reduced $\alpha $-width
($\theta _\alpha ^2$) for the members of the ground state bands of $^{90}$Sr,
$^{92}$Zr, $^{94}$Mo, $^{96}$Ru and $^{98}$Pd. The depths $V_0$ used to fit each
state at the corresponding experimental energy ($E_x$) are also indicated. See
the text for explanation about the channel radii. The experimental excitation
energies are from Refs.~\cite{B1997,B2012,AS2006,AS2008,SH2003}.}
\label{Table1}
\begin{ruledtabular}
\begin{tabular}{ccccccc}
&  &  &  &  &  &  \\[-2pt]
\multicolumn{7}{c}{$^{90}$Sr ($\alpha $-$^{86}$Kr system)} \\[2pt] \hline
&  &  &  &  &  &  \\[-8pt]
$J^\pi $ & $E_x$ (MeV) & $V_0$ (MeV) & $\langle R^2\rangle ^{1/2}$ (fm) & $%
\langle R^2\rangle ^{1/2}/(\langle r^2\rangle _\alpha ^{1/2}+\langle
r^2\rangle _{\mathrm{core}}^{1/2})$ & $\gamma _\alpha ^2$ (eV) &
 $\theta _\alpha ^2$ (10$^{-6}$) \\[2pt] \hline
&  &  &  &  &  &  \\[-6pt]
0$^{+}$ & 0.000 & 221.50 & 4.650 & 0.794 & 90.60 & 408.7 \\ 
2$^{+}$ & 0.832 & 220.36 & 4.652 & 0.794 & 95.05 & 428.8 \\ 
4$^{+}$ & 1.656 & 220.01 & 4.623 & 0.789 & 73.58 & 331.9 \\[2pt] \hline
&  &  &  &  &  &  \\ 
\multicolumn{7}{c}{$^{92}$Zr ($\alpha $-$^{88}$Sr system)} \\[2pt] \hline
&  &  &  &  &  &  \\[-8pt]
$J^\pi $ & $E_x$ (MeV) & $V_0$ (MeV) & $\langle R^2\rangle ^{1/2}$ (fm) & $%
\langle R^2\rangle ^{1/2}/(\langle r^2\rangle _\alpha ^{1/2}+\langle
r^2\rangle _{\mathrm{core}}^{1/2})$ & $\gamma _\alpha ^2$ (eV) &
 $\theta _\alpha ^2$ (10$^{-6}$) \\[2pt] \hline
&  &  &  &  &  &  \\[-6pt]
0$^{+}$ & 0.000 & 221.06 & 4.686 & 0.795 & 115.51 & 526.5 \\ 
2$^{+}$ & 0.935 & 219.62 & 4.695 & 0.796 & 126.89 & 578.3 \\ 
4$^{+}$ & 1.495 & 220.00 & 4.652 & 0.789 & 89.49 & 407.9 \\ 
6$^{+}$ & 2.957 & 218.64 & 4.610 & 0.782 & 61.56 & 280.6 \\ 
8$^{+}$ & 3.309 & 220.45 & 4.511 & 0.765 & 20.31 & 92.6 \\ 
10$^{+}$ & 4.297 & 220.67 & 4.426 & 0.751 & 5.57 & 25.4 \\ 
12$^{+}$ & 4.947 & 221.37 & 4.347 & 0.737 & 0.84 & 3.8 \\ 
14$^{+}$ & 6.046 & 220.11 & 4.310 & 0.731 & 0.07 & 0.3 \\[2pt] \hline
&  &  &  &  &  &  \\ 
\multicolumn{7}{c}{$^{94}$Mo ($\alpha $-$^{90}$Zr system)} \\[2pt] \hline
&  &  &  &  &  &  \\[-8pt]
$J^\pi $ & $E_x$ (MeV) & $V_0$ (MeV) & $\langle R^2\rangle ^{1/2}$ (fm) & $%
\langle R^2\rangle ^{1/2}/(\langle r^2\rangle _\alpha ^{1/2}+\langle
r^2\rangle _{\mathrm{core}}^{1/2})$ & $\gamma _\alpha ^2$ (eV) &
 $\theta _\alpha ^2$ (10$^{-6}$) \\[2pt] \hline
&  &  &  &  &  &  \\[-6pt]
0$^{+}$ & 0.000 & 221.80 & 5.109 & 0.859 & 708.39 & 3261.8 \\ 
2$^{+}$ & 0.871 & 220.37 & 5.118 & 0.861 & 763.03 & 3513.4 \\ 
4$^{+}$ & 1.574 & 219.98 & 5.087 & 0.856 & 615.02 & 2831.8 \\ 
6$^{+}$ & 2.423 & 219.70 & 5.033 & 0.847 & 399.44 & 1839.2 \\ 
8$^{+}$ & 2.956 & 220.54 & 4.943 & 0.831 & 178.00 & 819.6 \\ 
10$^{+}$ & 3.897 & 220.46 & 4.859 & 0.817 & 66.15 & 304.6 \\ 
12$^{+}$ & 4.192 & 221.78 & 4.765 & 0.801 & 14.48 & 66.7 \\[2pt] \hline
&  &  &  &  &  &  \\ 
\multicolumn{7}{c}{$^{96}$Ru ($\alpha $-$^{92}$Mo system)} \\[2pt] \hline
&  &  &  &  &  &  \\[-8pt]
$J^\pi $ & $E_x$ (MeV) & $V_0$ (MeV) & $\langle R^2\rangle ^{1/2}$ (fm) & $%
\langle R^2\rangle ^{1/2}/(\langle r^2\rangle _\alpha ^{1/2}+\langle
r^2\rangle _{\mathrm{core}}^{1/2})$ & $\gamma _\alpha ^2$ (eV) &
 $\theta _\alpha ^2$ (10$^{-6}$) \\[2pt] \hline
&  &  &  &  &  &  \\[-6pt]
0$^{+}$ & 0.000 & 221.72 & 5.103 & 0.852 & 595.69 & 2770.8 \\ 
2$^{+}$ & 0.833 & 220.38 & 5.112 & 0.853 & 636.52 & 2960.7 \\ 
4$^{+}$ & 1.518 & 220.02 & 5.081 & 0.848 & 509.85 & 2371.5 \\ 
6$^{+}$ & 2.150 & 220.25 & 5.020 & 0.838 & 310.25 & 1443.1 \\ 
8$^{+}$ & 2.950 & 220.36 & 4.943 & 0.825 & 149.95 & 697.5 \\ 
10$^{+}$ & 3.817 & 220.38 & 4.861 & 0.811 & 54.87 & 255.2 \\ 
12$^{+}$ & 4.418 & 220.92 & 4.776 & 0.797 & 13.11 & 61.0 \\ 
14$^{+}$ & 5.681 & 219.50 & 4.725 & 0.789 & 2.43 & 11.3 \\ 
16$^{+}$ & 6.442 & 218.44 & 4.702 & 0.785 & 0.20 & 0.9 \\[2pt] \hline
&  &  &  &  &  &  \\
\multicolumn{7}{c}{$^{98}$Pd ($\alpha $-$^{94}$Ru system)} \\[2pt] \hline
&  &  &  &  &  &  \\[-8pt]
$J^\pi $ & $E_x$ (MeV) & $V_0$ (MeV) & $\langle R^2\rangle ^{1/2}$ (fm) & $%
\langle R^2\rangle ^{1/2}/(\langle r^2\rangle _\alpha ^{1/2}+\langle
r^2\rangle _{\mathrm{core}}^{1/2})$ & $\gamma _\alpha ^2$ (eV) &
 $\theta _\alpha ^2$ (10$^{-6}$) \\[2pt] \hline
&  &  &  &  &  &  \\[-6pt]
0$^{+}$ & 0.000 & 221.78 & 5.101 & 0.842 & 497.31 & 2341.9 \\ 
2$^{+}$ & 0.863 & 220.32 & 5.114 & 0.844 & 541.49 & 2549.9 \\ 
4$^{+}$ & 1.541 & 220.00 & 5.082 & 0.839 & 430.80 & 2028.6 \\ 
6$^{+}$ & 2.112 & 220.34 & 5.019 & 0.829 & 256.86 & 1209.6 \\ 
8$^{+}$ & 2.773 & 220.74 & 4.941 & 0.816 & 119.26 & 561.6 \\ 
10$^{+}$ & 3.645 & 220.74 & 4.859 & 0.802 & 43.28 & 203.8 \\ 
12$^{+}$ & 4.447 & 220.76 & 4.782 & 0.790 & 10.90 & 51.3 \\ 
14$^{+}$ & 5.699 & 219.34 & 4.732 & 0.781 & 2.01 & 9.5 \\ 
16$^{+}$ & 6.749 & 217.60 & 4.715 & 0.778 & 0.17 & 0.8 
\end{tabular}
\end{ruledtabular}
\end{table*}

FIG.~\ref{Figure_Comp_94Mo} shows a comparison of different calculations for
the ground state band of $^{94}$Mo and the experimental spectrum. As
mentioned before, the $\alpha + \textrm{core}$ potential of shape
$(1 + \mathrm{Gaussian})\times \mathrm{W.S.}^{2}$ produces a rotational type spectrum
which provides only a rough description of the experimental energies. The
double folding potential used in Refs.~\cite{O1995,MOR1998} yields a good
description of the experimental spectrum between $0^{+}$ and $6^{+}$ by
introducing an $L$-dependent normalization factor. It should be taken into
account that the two potential forms mentioned above have been used
successfully to reproduce the $\alpha + ^{90}$Zr elastic scattering data.
The W.S.$+$W.S.$^{3}$ potential with fixed depth $V_0 = 220$ MeV describes
roughly the $0^{+}$ bandhead, but the introduction of a weak \mbox{$L$-dependence}
by Eq.~\eqref{eq:FuncPol_V0} provides a very good description of the
experimental band between the $0^{+}$ and $12^{+}$ states.

The root-mean-square (rms) intercluster separation is given by

\begin{equation}
\left\langle R^{2}\right\rangle_{G,J}^{1/2}=
\left[\int_{0}^{\infty}r^{2}\, u_{G,J}^{2}(r)\, dr\right]^{1/2}\;,\label{eq:rms}
\end{equation}

\noindent where $u_{G,J}(r)$ is the normalized radial wave function of a
$|G,J\rangle$ state. The value of $\langle R^2 \rangle ^{1/2}$ is seen to
decrease when one goes from the $0^{+}$ state to the highest spin state of each
band (see TABLE \ref{Table1}), except to $^{90}$Sr, where there are insufficient
known states for a complete analysis. This antistretching effect is found in
nuclei of other mass regions where the $\alpha $-cluster structure is studied,
considering different local potential forms \cite{BDV75,MRO88,O1995}. The same
effect is verified by Ohkubo \cite{O1995} for $^{94}$Mo through a double folding
potential.

The calculated rms intercluster separations have been used to estimate the rms
charge radius of the total nucleus by the equation

\begin{eqnarray}
\lefteqn{ \left\langle r^2\right\rangle _{T}=\frac{Z_\alpha }{Z_\alpha +
Z_{\text{core}}}\left\langle r^2\right\rangle _\alpha + \frac{Z_{\text{core}}}
{Z_\alpha +Z_{\text{core}}}\left\langle r^2\right\rangle _{\text{core}} {} }
\nonumber\\[4pt]
 & \qquad & {} + \frac{Z_\alpha A_{\text{core}}^2+Z_{\text{core}}A_\alpha ^2}
{(Z_\alpha +Z_{\text{core}})\left( A_\alpha +A_{\text{core}}\right) ^2}
\left\langle R^2\right\rangle \;,
\label{r2_T}
\end{eqnarray}

\noindent where $\langle r^2\rangle _{T} ^{1/2}$,
$\langle r^2\rangle _\alpha ^{1/2}$ and
$\langle r^2\rangle _{\text{core}} ^{1/2}$ are the rms charge
radii of the total system, the $\alpha$-cluster and the core, respectively.
Taking the experimental values \cite{A2004} of
$\langle r^2\rangle _\alpha ^{1/2}$ and
$\langle r^2\rangle _{\text{core}} ^{1/2}$ and considering
$\langle R^2 \rangle ^{1/2}$ as the calculated rms intercluster separation for
the ground state 0$^{+}$, we find the total rms radii in TABLE
\ref{Tab_Radii_rms_T}. Eq.~\eqref{r2_T} has not been applied for $^{98}$Pd,
since there are no experimental values for the rms charge radii of $^{94}$Ru and
$^{98}$Pd. TABLE \ref{Tab_Radii_rms_T} also shows the ratio of
$\langle r^{2}\rangle_{T}^{1/2}$ to the corresponding experimental rms charge
radius for each nucleus. The obtained ratios (very close to 1) indicate a high
level of agreement between the experimental data and the rms charge radii
predicted by the model.

\begin{table}
\caption{Rms charge radii $\langle r^{2}\rangle_{T}^{1/2}$ calculated
for the nuclei $^{90}$Sr, $^{92}$Zr, $^{94}$Mo and $^{96}$Ru
through eq.~\eqref{r2_T}. For each nucleus, it is shown
the ratio of $\langle r^{2}\rangle_{T}^{1/2}$ to the corresponding
experimental rms charge radius
($\langle r^{2}\rangle_{T\:\mathrm{exp}}^{1/2}$) \cite{A2004}.} 
\label{Tab_Radii_rms_T}
\begin{ruledtabular}
\begin{tabular}{ccc}
&  &  \\[-10pt]
Total nucleus & $\left\langle r^{2}\right\rangle_{T}^{1/2}$ (fm) &
 $\left\langle r^{2}\right\rangle_{T}^{1/2}/\left\langle r^{2}\right
\rangle_{T\:\mathrm{exp}}^{1/2}$\\[3pt] \hline
&  &  \\[-6pt] 
$^{90}$Sr & 4.220 & 0.990\\
$^{92}$Zr & 4.254 & 0.988\\
$^{94}$Mo & 4.322 & 0.993\\
$^{96}$Ru & 4.363 & 0.993\\
\end{tabular}
\end{ruledtabular}
\end{table}

Observing in TABLE \ref{Table1} the ratios

\begin{equation}
\frac{\langle R^2\rangle ^{1/2}}{\langle r^2\rangle _\alpha ^{1/2}+
\langle r^2\rangle _{\mathrm{core}}^{1/2}}
\label{RR}
\end{equation}

\noindent for each nucleus, where $\langle r^2\rangle _\alpha ^{1/2}$ and
$\langle r^2\rangle _{\mathrm{core}}^{1/2}$ are experimental values,
we verify a significant overlap between the $\alpha$-cluster and the core
for the ground states of the five nuclei. Because of the antistretching effect,
the overlap degree of the $\alpha + \textrm{core}$ system increases for higher
spin states, so that the $\alpha$-cluster character of the system is stronger
for the first states of the g.s.~bands.

A comparison of the same ratios indicates that the set of nuclei with
$G_{\mathrm{g.s.}}=16$ ($^{94}$Mo, $^{96}$Ru and $^{98}$Pd) has a more
pronounced $\alpha$-clustering than the set with \mbox{$G_{\mathrm{g.s.}}=14$}
($^{90}$Sr and $^{92}$Zr). Concerning the $0^{+}$ ground states of the five
nuclei, the ratios ${\scriptstyle
\langle R^2\rangle ^{1/2}/(\langle r^2\rangle _\alpha ^{1/2}+\langle
r^2\rangle _{\mathrm{core}}^{1/2})}$ for the $G_{\mathrm{g.s.}}=16$ nuclei
are approximately $6-8$ \% higher than the ratios for the 
$G_{\mathrm{g.s.}}=14$ nuclei. In this case, there is a clear effect of
the quantum number $G_{\mathrm{g.s.}}$, since the lower value of
$G_{\mathrm{g.s.}}$ results in a reduced value for the radial parameter $R$ and,
consequently, less extensive radial wave functions.

\begin{figure}[b]
\includegraphics[scale=0.35]{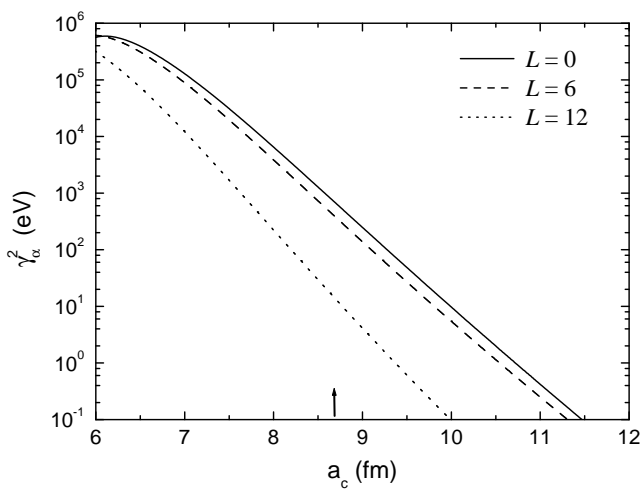}
\caption{Reduced $\alpha $-width ($\gamma _\alpha ^2$) as a function of the
channel radius ($a_c$) for the $0^{+}$, $6^{+}$ and $12^{+}$ states of $^{94}$Mo.
The arrow indicate the value of $a_c$ obtained by eq.~\eqref{Eq_ch_radius}.
The $\gamma _\alpha ^2$ axis is represented in a log$_{10}$ scale.}
\label{Figure_Red_Width_94Mo}
\end{figure}

The radial wave functions are also used for the calculation of the reduced
$\alpha $-width \cite{AY1974,MSV1970}

\begin{equation}
\gamma _{\alpha}^2=\left( \frac{\hbar^2}{2\mu a_{c}}\right) u^2(a_c)\left[
\int_0^{a_c}|u(r)|^2dr\right] ^{-1}\;,
\end{equation}

\noindent where $\mu $ is the reduced mass of the system, $u(r)$ is the radial
wave function of the state and $a_c$ is the channel radius. In general,
$a_c$ lies near the radius of the Coulomb barrier top. It is known that the
$\gamma _\alpha ^2$ value is very sensitive to the choice of the channel radius.
As an example, FIG.~\ref{Figure_Red_Width_94Mo} shows $\gamma _\alpha ^2$ as a
function of $a_c$ for the $0^{+}$, $6^{+}$ and $12^{+}$ states of $^{94}$Mo;
it is verified that the decrease of $\gamma _\alpha ^2$ with increasing $a_c$
is of exponential type (and linear type with the log$_{10}$ scale for
$\gamma _\alpha ^2$) from $\approx 7$ fm.

\begin{figure}
\includegraphics[scale=0.35]{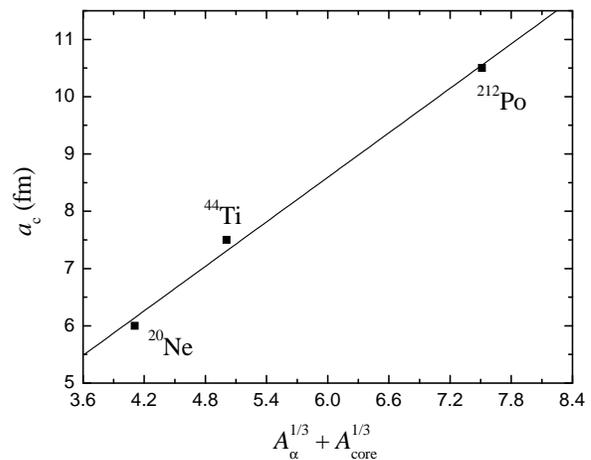}
\caption{Channel radii employed in different works for the
$\alpha + \textrm{core}$ systems in $^{20}$Ne \cite{AY1974,FHI1980}, $^{44}$Ti
\cite{MRO88} and $^{212}$Po \cite{O1995} as a function of
\mbox{$A_{\alpha}^{1/3} + A_{\mathrm{core}}^{1/3}$}. The full line shows the
linear function fitted for the three radii. See the text for details of the
linear function.}
\label{Figure_Ch_Radii}
\end{figure}

In this work, a procedure that avoids an arbitrary choice of channel radius
is used. The value of $a_c$ is given by the relation

\begin{equation}
a_c = 1.295(A_{\alpha}^{1/3} + A_{\mathrm{core}}^{1/3}) + 0.824\;,
\label{Eq_ch_radius}
\end{equation}

\noindent obtained from a linear fit (see FIG.~\ref{Figure_Ch_Radii}) that
considers other channel radii used for different $\alpha + \textrm{core}$
systems in the literature \cite{O1995,MRO88,FHI1980}. The dimensionless reduced
$\alpha$-width $\theta _\alpha ^2$ is defined as the ratio of
$\gamma _{\alpha}^2$ to the Wigner limit, that is,

\begin{equation}
\theta _\alpha ^2=\frac{2\mu a_{c}^2}{3\hbar ^2}\gamma _{\alpha}^2\;. 
\end{equation}

\noindent Qualitatively, a large value of $\theta _\alpha ^2$ ($\approx 1$) is
interpreted as an evidence of a high degree of $\alpha$-clustering.

The g.s.~bands of $^{92}$Zr, $^{94}$Mo, $^{96}$Ru and $^{98}$Pd show a
rapid decrease of $\gamma _\alpha ^2$ with the increasing spin.
In agreement with the analysis of the rms intercluster separations, the
behavior of $\gamma _\alpha ^2$ suggests a stronger $\alpha $-cluster
character for the first members of these bands. For the five nuclei, the
dimensionless reduced $\alpha $-widths $\theta _\alpha ^2$ present a very small
fraction of the Wigner limit, even for the first members of the band, indicating
a general weak $\alpha $-cluster character for the g.s.~bands.

The effect of the quantum number $G$ on $\theta _\alpha ^2$ is analogous to that
observed for the ratio of eq.~\eqref{RR}. The set of nuclei with
$G_{\mathrm{g.s.}}=16$ present $\theta _\alpha ^2$ values that are approximately
one order of magnitude higher than the corresponding $\theta _\alpha ^2$ values
for the set with $G_{\mathrm{g.s.}}=14$. Therefore, in addition to the results
obtained for the rms radii, it is indicated that the set with
$G_{\mathrm{g.s.}}=16$ has a higher $\alpha$-cluster degree in comparison
with the $G_{\mathrm{g.s.}}=14$ set.

Comparing the $J^{\pi}$ states of the five nuclei in TABLE \ref{Table1}, it is
noted that $^{94}$Mo has the highest $\theta _\alpha ^2$ values. This result
corroborates the consideration of $^{94}$Mo as the preferential nucleus for
$\alpha $-clustering in its mass region.

The model also allows the calculation of the $B(E2)$ transition rates
between the states of an $\alpha$-cluster band. In the case where the cluster
and the core have zero spins, this quantity is given by

\begin{eqnarray}
\lefteqn{B\left(E2;G,J\rightarrow J-2\right)= {} }
                     \nonumber\\[4pt]
& & {} 
\frac{15}{8\pi}\,\beta_{2}^{2}\,\frac{J\left(J-1\right)}
{\left(2J+1\right)\left(2J-1\right)}\left\langle
 r_{J,J-2}^{2}\right\rangle ^{2}\;,
\label{BE2_a}
\end{eqnarray}

\noindent where

\begin{equation}
\left\langle r_{J,J-2}^{2}\right\rangle =
\int_{0}^{\infty}r^{2}\, u_{G,J}(r)\, u_{G,J-2}(r)dr \;, \label{BE2_b}
\end{equation}

\noindent $\beta_{2}$ is the recoil factor, given by

\begin{equation}
\beta_{2}=\frac{Z_{\alpha}A_{\mathrm{core}}^{2}+Z_{\mathrm{core}}A_{\alpha}^{2}}
{\left(A_{\alpha}+A_{\mathrm{core}}\right)^{2}}\;,
\label{BE2_c}
\end{equation}

\noindent $u_{G,J}(r)$ and $u_{G,J-2}(r)$ are the radial wave functions of the
initial $|G,J\rangle$ state and final $|G,J-2\rangle$ state, respectively.

The calculated $B(E2)$ transition rates for the ground state bands are presented
in TABLE \ref{Table_BE2}. A comparison of the calculated values for the five
nuclei and the experimental data shows that the model can reproduce, with few
exceptions, the correct order of magnitude of the experimental $B(E2)$ values
without the use of effective charges. These results may be considered
satisfactory since, in shell-model calculations for medium-heavy nuclei
\cite{ZWG1999,LPF2000}, large effective charges are necessary to reproduce the order of
magnitude of the experimental data.

\begin{table}
\caption{Comparison of the calculated $B(E2)$ transition rates for the ground
state bands of $^{90}$Sr, $^{92}$Zr, $^{94}$Mo, $^{96}$Ru and $^{98}$Pd with the
corresponding experimental data \cite{B1997,B2012,AS2006,AS2008,SH2003}. The
calculated values have been obtained without effective charges.}
\label{Table_BE2}
\begin{ruledtabular}
\begin{tabular}{ccc}
&  &  \\[-2pt]
\multicolumn{3}{c}{$^{90}$Sr ($\alpha $-$^{86}$Kr system)} \\[2pt] \hline
&  &  \\[-7pt]
$J^\pi $ & $B(E2;J\rightarrow J-2)$ (W.u.) &
 $B(E2)_{\mathrm{exp.}}$ (W.u.) \\[2pt] \hline
&  &  \\[-6pt]
2$^{+}$ & 5.569 & 8.5(24) \\
4$^{+}$ & 7.635 & 5.2(9) \\[2pt] \hline
&  &  \\
\multicolumn{3}{c}{$^{92}$Zr ($\alpha $-$^{88}$Sr system)} \\[2pt] \hline
&  &  \\[-7pt]
$J^\pi $ & $B(E2;J\rightarrow J-2)$ (W.u.) &
 $B(E2)_{\mathrm{exp.}}$ (W.u.) \\[2pt] \hline
&  &  \\[-6pt]
2$^{+}$ & 5.623 & 6.4(6) \\
4$^{+}$ & 7.661 & 4.05(12) \\ 
6$^{+}$ & 7.727 & $\geq$ 0.00098 \\ 
8$^{+}$ & 6.610 & 3.59(22) \\ 
10$^{+}$ & 5.268 &  \\ 
12$^{+}$ & 3.569 & $\geq$ 0.056 \\ 
14$^{+}$ & 1.859 &  \\[2pt] \hline
&  &  \\ 
\multicolumn{3}{c}{$^{94}$Mo ($\alpha $-$^{90}$Zr system)} \\[2pt] \hline
&  &  \\[-7pt]
$J^\pi $ & $B(E2;J\rightarrow J-2)$ (W.u.) &
 $B(E2)_{\mathrm{exp.}}$ (W.u.) \\[2pt] \hline
&  &  \\[-6pt]
2$^{+}$ & 7.757 & 16.0(4) \\
4$^{+}$ & 10.738 & 26(4) \\ 
6$^{+}$ & 10.898 &  \\ 
8$^{+}$ & 9.798 & 0.0049(8) \\ 
10$^{+}$ & 8.303 &  \\ 
12$^{+}$ & 6.237 &  \\[2pt] \hline
&  &  \\ 
\multicolumn{3}{c}{$^{96}$Ru ($\alpha $-$^{92}$Mo system)} \\[2pt] \hline
&  &  \\[-7pt]
$J^\pi $ & $B(E2;J\rightarrow J-2)$ (W.u.) &
 $B(E2)_{\mathrm{exp.}}$ (W.u.) \\[2pt] \hline
&  &  \\[-6pt]
2$^{+}$ & 7.539 & 18.4(4) \\
4$^{+}$ & 10.430 & 20.7(15) \\ 
6$^{+}$ & 10.512 & 14(5) \\ 
8$^{+}$ & 9.584 & 6.0(22) \\ 
10$^{+}$ & 8.098 & 12.7(15) \\ 
12$^{+}$ & 6.190 & 13.1(19) \\ 
14$^{+}$ & 4.341 & 2.63(23) \\ 
16$^{+}$ & 2.186 & $>$ 11 \\[2pt] \hline
&  &  \\ 
\multicolumn{3}{c}{$^{98}$Pd ($\alpha $-$^{94}$Ru system)} \\[2pt] \hline
&  &  \\[-7pt]
$J^\pi $ & $B(E2;J\rightarrow J-2)$ (W.u.) &
 $B(E2)_{\mathrm{exp.}}$ (W.u.) \\[2pt] \hline
&  &  \\[-6pt]
2$^{+}$ & 7.364 &  \\
4$^{+}$ & 10.195 & \\ 
6$^{+}$ & 10.253 & \\ 
8$^{+}$ & 9.315 & (without exp. \\ 
10$^{+}$ & 7.882 & data) \\ 
12$^{+}$ & 6.097 & \\ 
14$^{+}$ & 4.257 & \\ 
16$^{+}$ & 2.179 & 
\end{tabular}
\end{ruledtabular}
\end{table}

Observing the $2^{+} \rightarrow 0^{+}$ and $4^{+} \rightarrow 2^{+}$
transitions for the nuclei $^{90}$Sr, $^{92}$Zr, $^{94}$Mo and $^{96}$Ru, it is
noted that the experimental $B(E2)$ rates for $^{94}$Mo and $^{96}$Ru are
\mbox{$\approx 2-6 \, \times$} higher than the experimental rates for $^{90}$Sr
and $^{92}$Zr. This fact indicates that the application of the quantum numbers
$G_{\mathrm{g.s.}}=14$ for $^{90}$Sr and $^{92}$Zr, and $G_{\mathrm{g.s.}}=16$
for $^{94}$Mo and $^{96}$Ru, is favourable for a better description of the
experimental transition rates. The functions $u_{G,J}(r)$ with
$G_{\mathrm{g.s.}}=16$ are more extensive at the surface region than the
functions $u_{G,J}(r)$ with $G_{\mathrm{g.s.}}=14$, which results in higher
values for the calculated $B(E2)$ rates of $^{94}$Mo and $^{96}$Ru.

The $B(E2)$ values calculated by Mohr \cite{M2008} with a double folding nuclear potential,
also without effective charges, are $\approx 1.3-1.4 \, \times$ higher than the values of this
work for $^{90}$Sr and $^{92}$Zr, and $\approx 1.15 \, \times$ higher for $^{94}$Mo and $^{96}$Ru.
Compared to Ref.~\cite{M2008}, the present work produces results closer to the experimental values
only for $^{90}$Sr ($4^{+} \rightarrow 2^{+}$ transition) and $^{92}$Zr. In general, it can be
stated roughly that the two works produce similar results in comparison with the experimental
$B(E2)$ rates.

\subsection{Negative parity bands}

The comparison of theoretical negative parity bands with the experimental
spectra is more difficult because of the scarce experimental negative parity
levels with definite assignments. Previous works \cite{O1995,M2008,SM2005}
discuss the existence of an $\alpha + \textrm{core}$ negative parity band for
$^{94}$Mo and indicate that the 1$^{-}$ bandhead should lie at $E_x = 6-7$ MeV;
however, such predictions are not compared with experimental energies of
$^{94}$Mo. Therefore, at this stage, any calculation concerning the negative
parity bands of the studied nuclei should be interpreted as a theoretical
prediction that lacks complementary experimental information.

\begin{figure*}
\includegraphics[scale=0.76]{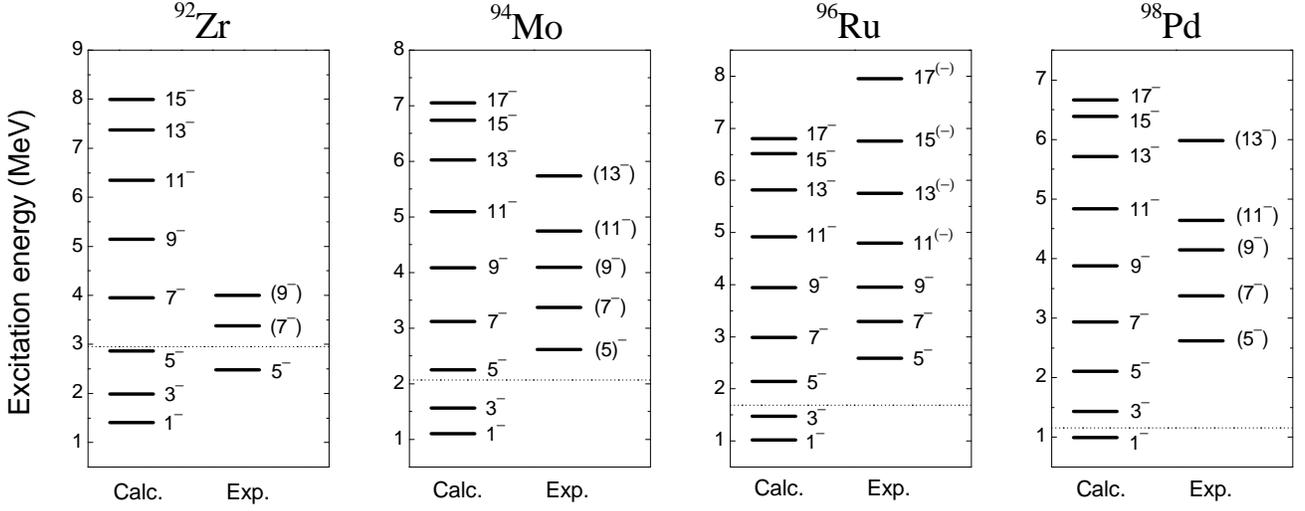}
\caption{Calculated $\alpha + \textrm{core}$ energies for the negative parity
bands of $^{92}$Zr ($G=15$), $^{94}$Mo ($G=17$), $^{96}$Ru ($G=17$) and
$^{98}$Pd ($G=17$) in comparison with the available experimental levels.
The dotted lines indicate the $\alpha + \textrm{core}$ thresholds.
A depth $V_0 = 238$ MeV was applied. The experimental data are from
Refs.~\cite{B2012,AS2006,AS2008,SH2003}.}
\label{Figure_Espec_Neg}
\end{figure*}

In this work, we have selected the lowest negative parity levels connected by
$E2$ transitions (or possible $E2$ transitions) for a comparison with the
calculated negative parity bands, which are shown in
FIG.~\ref{Figure_Espec_Neg}. The quantum number $G = 15$ is applied to $^{92}$Zr
and $G = 17$ is applied to $^{94}$Mo, $^{96}$Ru and $^{98}$Pd. The experimental
spectrum of $^{90}$Sr \cite{B1997} presents insufficient data concerning
negative parity levels and a comparison with the theoretical negative parity
band is unfeasible. The values for the parameters $a$, $b$ and $R$ are the same
used for the calculation of the g.s.~bands, while a depth $V_0 = 238$ MeV is
applied to the negative parity bands of the four nuclei. The calculated bands
provide a good description of the spacings between most of the experimental
energy levels. In the case of the experimental $15^{(-)}$ and $17^{(-)}$ states
of $^{96}$Ru, the energy spacing is roughly described because of the compression
of the theoretical band at the higher spin states. The individual analysis
of the four negative parity bands shows that $V_0 = 237$ MeV and $V_0 = 239$ MeV
are more appropriate depths to describe the experimental spectra of $^{98}$Pd
and $^{92}$Zr, respectively.

Analysing the calculated negative parity states above the
$\alpha + \textrm{core}$ threshold, it is verified that the extension of the
effective potential barrier varies within the range
$107 \; \mathrm{fm} < \Delta r < 635 \; \mathrm{fm}$ for the resonant states
nearest to the threshold, and within the range
$25 \; \mathrm{fm} < \Delta r < 29 \; \mathrm{fm}$ for the highest spin states
($15^{-}$ for $^{92}$Zr and $17^{-}$ for $^{94}$Mo, $^{96}$Ru and $^{98}$Pd).
The height of the effective potential barrier in relation to the resonant
energies varies within the range
$13 \; \mathrm{MeV} < \Delta E < 17 \; \mathrm{MeV}$ for the resonant states
nearest to the threshold, and within the range
$28 \; \mathrm{MeV} < \Delta E < 32 \; \mathrm{MeV}$ for the highest spin
states. Therefore, the negative parity bands repeat the feature of the ground
state bands in which the states above the threshold are practically bound.

Regarding the experimental energy levels, the three negative parity states of
$^{92}$Zr are populated in the \mbox{$^{88}$Sr($^7$Li,$2np\gamma$)$^{92}$Zr}
reaction \cite{B2012,B1976}, therefore, being favorable candidates for an
association with the calculated $\alpha $-cluster states. There is no
information from $\alpha$-transfer reactions for the selected negative parity
states of $^{94}$Mo, $^{96}$Ru and $^{98}$Pd \cite{AS2006,AS2008,SH2003}. 

The depth $V_0 = 238$ MeV is $\approx 8.2$ \% higher than the depth
$V_0 = 220$ MeV used to describe the g.s.~bands, indicating a moderate parity
dependence for the $\alpha + \textrm{core}$ potential. The variation of the
depth parameter for different bands is also observed in other
$\alpha + \textrm{core}$ systems \cite{BDV75,BMP1999_NPA652,MAB1983}.
E.g., a calculation of the $\alpha$-cluster structure in light nuclei
\cite{BDV75} using a folding nuclear potential with a depth parameter
$\overline{f}$ shows that the values of $\overline{f}$ for the negative parity
bands of $^{16}$O and $^{20}$Ne ($G = 9$) are, respectively, $\approx 8.8$ \%
and $\approx 7.1$ \% higher than the values of the same parameter for the
corresponding positive parity bands ($G = 8$).

Because of the enhanced value of $V_0$, the bandheads of the theoretical
negative parity bands lie close to \mbox{$E_x = 1$ MeV}, which differs from
previous predictions for $^{94}$Mo \cite{O1995,M2008,SM2005}.
According to the criterion adopted to select the experimental levels,
there are no experimental $1^{-}$ and $3^{-}$ states for comparison
with the calculated bands (see FIG.~\ref{Figure_Espec_Neg}).
An experimental $3^{-}$ state at $E_x = 2.340$ MeV is present in the spectrum of
$^{92}$Zr, however, such state is not connected with the first $5^{-}$ state
($E_x = 2.486$ MeV) by $\gamma$-transition. The same condition occurs for the
experimental $3^{-}$ ($E_x = 2.534$ MeV) and $(5)^{-}$ ($E_x = 2.611$ MeV)
states of $^{94}$Mo. It is possible that the theoretical $1^{-}$ states are
experimentally nonexistent near the predicted energies, since the
spectra measured by several reactions for the studied nuclei  
\cite{B2012,AS2006,AS2008,SH2003} do not indicate the presence of a $1^{-}$
state around $E_x = 1$ MeV. However, the low energy location of the 
calculated negative parity bands is mainly justified by the presence of the 
$5^{-}$, ($7^{-}$) and ($9^{-}$) experimental levels of $^{92}$Zr from
$E_x \approx 2.5$ MeV, which are populated in a $\alpha$-transfer reaction. 
Additionally, the similar features of the spectra favour a unified description
of the negative parity bands of the four nuclei. Further experimental data
are essential for a better evaluation of the calculations described in this
section.

\section{Relation with $\alpha$+core optical potentials}

The $\alpha + \textrm{core}$ potential may be employed as the real part of
an optical potential for the analysis of $\alpha$ elastic scattering. However,
is often necessary to readjust one or more parameters of the potential for a
satisfactory description of the experimental $\alpha$-scattering data.
The adjustment of the parameters of the imaginary part of the optical potential is also
important in this procedure. The works of Buck {\it et al.}~\cite{BJM1995,BJM1996} show that
the W.S.$+$W.S.$^{3}$ potential may be used successfully to describe the data of
$\alpha$ elastic scattering on nuclei $^{16}$O, $^{40}$Ca and $^{208}$Pb. Therefore, it is
interesting to verify if the $\alpha + \textrm{core}$ potentials discussed in this work can
be applied in the same way.

Two important quantities used for describing an optical potential are the volume integral
per nucleon pair

\begin{equation}
J_{R}=\frac{4\pi}{A_{\alpha} \, A_{\mathrm{core}}}
\int_{0}^{\infty}\, V_{N}(r)\,r^{2}\, dr \;,\label{eq:int_vol}
\end{equation}

\noindent and the root-mean-square (rms) radius associated with the potencial

\begin{equation}
r_{\mathrm{rms},R} = \left[
\frac{\displaystyle \int_{0}^{\infty}\, V_{N}(r)\,r^{4}\, dr}
{\displaystyle \int_{0}^{\infty}\, V_{N}(r)\,r^{2}\, dr} \right]^{1/2}\;.
\label{eq:rms_pot}
\end{equation}

\noindent Eqs.~\eqref{eq:int_vol} and \eqref{eq:rms_pot} refer specifically to the nuclear
real part of the optical potential.

\begin{table}
\caption{Calculated values of the volume integral per nucleon pair
($J_R$) and root-mean-square radius ($r_{\mathrm{rms},R}$) for the
$\alpha + \textrm{core}$ nuclear potentials employed in this work.}
\label{Tab_Int_vol}
\begin{ruledtabular}
\begin{tabular}{cccc}
Nucleus & $V_0$ (MeV) & $J_V$ (MeV fm$^3$) & $r_{\mathrm{rms},R}$ (fm)\\[2pt] \hline
&  &  &  \\[-6pt] 
$^{90}$Sr & 220 & 254.1 & 4.697 \\
$^{90}$Sr & 238 & 274.8 & 4.697 \\
$^{92}$Zr & 220 & 244.8 & 4.682 \\
$^{92}$Zr & 238 & 264.8 & 4.682 \\
$^{94}$Mo & 220 & 310.5 & 4.974 \\
$^{94}$Mo & 238 & 335.9 & 4.974 \\
$^{96}$Ru & 220 & 307.5 & 4.988 \\
$^{96}$Ru & 238 & 332.6 & 4.988 \\
$^{98}$Pd & 220 & 303.6 & 4.998 \\
$^{98}$Pd & 238 & 328.4 & 4.998 \\
\end{tabular}
\end{ruledtabular}
\end{table}

TABLE \ref{Tab_Int_vol} shows the $J_R$ and $r_{\mathrm{rms},R}$ values calculated for the
$\alpha + \textrm{core}$ potentials of this work, with the depths applied to the
positive ($V_0 = 220$ MeV) and negative ($V_0 = 238$ MeV) parity bands. As usual,
the negative sign of $J_R$ is omitted.

Ref.~\cite{AMA1996} analyzes the $\alpha$ elastic scattering on $^{90}$Zr at the energies
$E_{\alpha,\mathrm{lab}} =$ 31.0, 59.1, 104.0 and 141.7 MeV using a double folding nuclear
potential with DDM3Y interaction; for these energies, the values $J_R \approx 270-335$ \mbox{MeV fm$^3$}
and $r_{\mathrm{rms},R} \approx 5.0$ fm have been obtained. The $J_R$ values obtained in
the present work for $^{94}$Mo are within the range mentioned previously, and
the radius $r_{\mathrm{rms},R}$ obtained in the present work is in excellent agreement with those from
Ref.~\cite{AMA1996}. The works of Ohkubo \cite{O1995} and Michel, Reidemeister and Ohkubo \cite{MRO2000},
which use different forms of $\alpha + ^{90}$Zr nuclear potential to describe the $\alpha$
elastic scattering data and energy levels, find $J_R$ values very close to the ones
obtained in this work for $^{94}$Mo.

Refs.~\cite{MKF2013,FGM2001} analyze the $\alpha$ elastic scattering on $^{92}$Mo at the energies
$E_{\mathrm{c.m.}} \approx$ 13, 16 and 19 MeV, also using a double folding nuclear potential
with DDM3Y interaction and different parametrizations for the imaginary part of the
optical potential. The two references show volume integral values
in the range $J_R \approx 320-360$ MeV fm$^3$ and rms radii in the range
$r_{\mathrm{rms},R} \approx 4.9-5.1$ fm. The $J_R$ value obtained in this work for $^{96}$Ru with
$V_0 = 220$ MeV is very close to the range mentioned before, while the $J_R$ value
with $V_0 =$ 238 MeV is within the same range. The calculated value of $r_{\mathrm{rms},R}$ for $^{96}$Ru
is also in excellent agreement with those mentioned in Refs.~\cite{MKF2013,FGM2001}.

Refs.~\cite{AMA1996, MKF2013} indicate that there is a systematic behavior of $J_R$ for
$\alpha$ elastic scattering in different mass regions, resulting in values in the range
$J_R \approx 320-370$ \mbox{MeV fm$^3$} for almost all analyzed energies around the Coulomb barrier.
Therefore, the $J_R$ values calculated in the present work for $^{90}$Sr and $^{92}$Zr are
below the mentioned range. This demonstrates that the choice of the quantum number $G$ can strongly change
the $J_R$ value for a certain $\alpha + \textrm{core}$ potential. However, one can not state in
advance that the $\alpha + ^{86}$Kr and $\alpha + ^{88}$Sr potentials are \mbox{unsuitable} for description of
$\alpha$ elastic scattering at low energies, since the reproduction of the angular distributions
also depends on the fit of the parameters of the imaginary part of the optical potential.
E.g., the work of Buck {\it et al.}~\cite{BJM1996} states that it is possible to describe the data from
$\alpha$ elastic scattering on $^{208}$Pb at low energies with different fits of optical potential
parameters, and such fits are related to different quantum numbers $G_{\mathrm{g.s.}}$.

As $^{94}$Ru is an unstable nucleus, the values of $J_R$ and $r_{\mathrm{rms},R}$ for the
$\alpha + ^{94}$Ru system are not compared with other references. The use of the $\alpha + \textrm{core}$
potentials of this work for analysis of $\alpha$ elastic scattering data is a project
to be performed in a next stage.

\section{Conclusions}

The $\alpha$-cluster structure in the nuclei $^{90}$Sr, $^{92}$Zr, $^{94}$Mo,
$^{96}$Ru and $^{98}$Pd was investigated through a local $\alpha+\textrm{core}$
potential. A good account of the experimental ground state bands of the five
nuclei is given with only one variable radial parameter $R$.
Such parameter has approximately a linear relation with the
total nuclear radius and with the sum of the $\alpha$-cluster and core radii.
An analysis of the depth parameter $V_0$ shows a weak dependence on the angular
momentum $L$, described satisfactorily by a polynomial function
$\tilde{V}_0(L)$. The rms inter-cluster separations and reduced $\alpha$-widths
calculated for the ground state bands indicate a decrease of the
$\alpha$-cluster intensity with the increasing spin, and that the nuclei
associated with the quantum number $G_{\mathrm{g.s.}}=16$ ($^{94}$Mo, $^{96}$Ru
and $^{98}$Pd) have a significantly higher degree of $\alpha$-clustering than
those with $G_{\mathrm{g.s.}}=14$ ($^{90}$Sr and $^{92}$Zr). The $B(E2)$
transition rates calculated for the ground state bands reproduce the order of
magnitude of almost all experimental data without the use of effective charges.
It is verified that the employment of $G_{\mathrm{g.s.}}=14$ for $^{90}$Sr and
$^{92}$Zr and $G_{\mathrm{g.s.}}=16$ for $^{94}$Mo and $^{96}$Ru is favourable
for a better description of the experimental transition rates.

The calculated negative parity bands provide a good account of the spacings
between most of the experimental levels, considering that only the depth
parameter was changed from the value employed for the ground state
bands; this change indicates a moderate parity dependence for the
$\alpha+\textrm{core}$ potential. For the moment, the calculations for the negative
parity bands should be interpreted as an exploratory attempt to describe the
available experimental data, since many theoretical levels could not be compared
to experimental levels or were associated with undefined experimental assignments.

The $\alpha+\textrm{core}$ potentials applied in this work were used to calculate the
volume integral per nucleon pair and rms radius. The $J_R$ values calculated for
the $\alpha + ^{90}$Zr and $\alpha + ^{92}$Mo potentials, referring to the positive and
negative parity bands, are compatible or very close to those obtained from the respective
optical potentials of $\alpha$ elastic scattering at several energies \cite{AMA1996,MKF2013,FGM2001}.
The calculated rms radii are in excellent agreement with those obtained from the same optical
potentials. These results give a strong indication that the $\alpha + ^{90}$Zr and $\alpha + ^{92}$Mo
potentials of this work may be used to describe $\alpha$ elastic scattering data. Due to the
lower $G_{\mathrm{g.s.}}$ number for $^{90}$Sr and $^{92}$Zr, the values of $J_R$ for the
$\alpha + ^{86}$Kr and $\alpha + ^{88}$Sr potentials are slightly below the range expected
for $\alpha$-scattering energies around the Coulomb barrier.

The main conclusion of the present work is that the even-even $N = 52$ nuclei around
$^{94}$Mo present similar $\alpha + \mathrm{core}$ structures. In addition, it
is pointed that the $\alpha$-cluster model can be extended successfully for
other nuclei of the intermediate mass region.
New experimental data, mainly from $\alpha$-transfer reactions, will be very
important for a better evaluation of the calculated $\alpha$-cluster states.
The Local Potential Model should be used in a next work for investigating the
$\alpha$-cluster structure in other nuclei of the same mass region.
Another project to be performed is the application of the $\alpha + \mathrm{core}$
potentials of this work for analysis of $\alpha$ elastic scattering data.

\begin{acknowledgments}

The authors thank the members of the Nuclear Spectroscopy with Light Ions Group
of IFUSP for the discussions on this work. We used resources of LCCA -
Laboratory of Advanced Scientific Computation of University of S\~{a}o Paulo.
This work was financially supported by CAPES.

\end{acknowledgments}

\end{document}